\documentstyle[12pt,a4,psfig]{article}

\begin{document}
\thispagestyle{empty}
\hskip 10.60truecm FUB-HEP/95-16
\vspace*{2.truecm}
\begin{center}
{\LARGE {\bf New results of \\
\vspace{0.3truecm}
the Berliner relativistic quark model$^{\S ~*}$}}\\ 
\end{center}
\vspace{2em}
\begin{center}
\large
 C.Boros\\
\vspace{2em}
{\it Institut f\"ur Theoretische Physik der Freien Universit\"at Berlin,\\
Arnimallee 14, 14195 Berlin, Germany}
\end{center}

\vspace*{1.5truecm}
 
\begin{center}
{\bf Abstract}
\end{center}
 
\vspace*{0.2truecm}

\begin{parbox}{12.cm}

It is pointed out that the recently measured 
left-right asymmetries in inclusive pion and lambda hyperon production 
processes are in very good agreement with the relativistic quark model, 
proposed some time ago. 
Further predictions based on this model for  hyperon productions,  
for lepton pair productions and for W/Z-boson productions  
are also presented. 
\end{parbox}

\vspace*{\fill}

\vskip 2.0truecm

\noindent
------------------------------------------------

$\S$Invited talk given at the VI International Workshop on 
Spin Phenomena in High Energy Physics, Protvino, Russia September 18-23.

$^*$Supported in part by Deutsche Forschungsgemeinschaft
DFG: Me 470/7-1.

\newpage
\vspace*{2.truecm}

\section{Introduction}
Inclusive pion, and $\eta$-meson production experiments 
using transversely polarized 200 $GeV/c$ proton or 
antiproton projectiles and unpolarized targets has shown that
the left-right asymmetries are not only very large in the 
 fragmentation region of the polarized hadrons 
but also that they depend  on the flavor content of the polarized 
colliding objects and on that of the  produced particles 
\cite{adams,yoko}.  
Recently also the left-right asymmetry in single-spin $\Lambda$ hyperon 
production in proton-proton collisions  
has been measured \cite{bravar}. 
The data shows that the asymmetry is zero for small 
$x_F$ ($x_F<0.2$), than slightly positive in the region $0.2<x_F<0.5$ 
(although it is compatible with zero within the error bars in this
region) and above $x_F=0.5$ it is large and negative. 
An interesting  feature of the observed asymmetry is 
that it becomes significant for 
much larger $x_F$ than that for pion production.

These asymmetries are expected to vanish in the usual 
leading twist pQCD-based hard-scattering model \cite{kane}, suggesting 
that soft dynamics should be important in understanding 
the experimental data. Several attempts have been made in trying to 
include higher twist effects in the framework of the hard-scattering 
model \cite{qcd}. 
In this talk I want to focus on a somewhat different approach which 
has been proposed by the Berliner group and is sometimes called 
the Berliner relativistic quark model. 
These phenomenological model has been  successfully used in describing 
the existing experimental data 
and predicting the outcome of some single-spin experiments long before the 
corresponding data were available [6-9]. 
In the following I wish to point out that one can also understand the 
$\Lambda$ hyperon data in terms of the  proposed model and  
present predictions for other hyperon production processes \cite{lambda}.  
It has been emphasized  that as a consequence 
of the model not only meson and hyperon production processes 
but also the  corresponding lepton pair  and 
W/Z-boson  production  should show significant left-right asymmetries 
and by measuring these asymmetries one can test the proposed  model 
\cite{lett,brief}.   
Thus it is crucial to have definite predictions for these 
processes. In this talk 
I  discuss the production of Drell-Yan
pairs using pion beam and different polarized targets and the 
asymmetries involved in single-spin inclusive 
W/Z-boson production in proton-proton and antiproton-proton 
collisions. 
The material presented here has been obtained 
in collaboration with Prof. Meng and Dr. Liang. 

\section{The physical picture}
Let us summarize first briefly the key points of the model [6-9]. 

(i) Valence quarks in hadron are treated as relativistic dirac particles
in an effective confining potential. 
It can be readily shown that orbital motion is always involved. 
Furthermore the direction of the orbital motion is determined 
uniquely by the polarization of the valence-quark. 

(ii) Valence quarks in a polarized hadron are polarized either in the
same or in the opposite direction as the hadron. 
This is determined by the wave function of the hadron. For proton 
we have that 5/3 of the u-quarks are polarized in the same and 1/3 of
them in the opposite direction as the polarized hadron. 
For d-quarks we have 1/3 and 2/3 respectively. 

(iii) In the fragmentation regions of the high-energy hadron-hadron
collisions a significant part of the mesons/hyperons are products of
direct-formation 
processes between the valence-quarks of the one hadron with suitable
sea-quarks of the other hadron. 

(iv) Because of the fact, that hadrons are extended objects and their
constituents interact with each other we expect that there exists a 
surface effect, i.e. only those mesons, hyperons which are produced on
the  front surface  of the polarized hadron, retain information about
the polarization. 

In order to compare the proposed picture with the recent experiments,
let us recall how the left-right asymmetry is defined for the 
reaction $p(\uparrow) + p(0) \rightarrow (pp) +X$ where 
$pp$ stands for the produced particle, which is 
either a meson or a baryon or  a lepton pair.  
\begin{equation}
A_N(x_F,Q|s)\equiv \frac{N(x_F,Q|s,\uparrow)-N(x_F,Q|s,\downarrow)}
{N(x_F,Q|s,\uparrow)+N(x_F,Q|s,\downarrow)}
\end{equation}
where 
\begin{equation}
N(x_F,Q|s,\uparrow /\downarrow)\equiv \frac{1}{\sigma_{in}}
 \int_D d^2p_{\perp} \frac{d\sigma}{dx_F dp^2_{\perp}(x_F,\vec{p_{\perp}},Q|s)}
   (\uparrow /\downarrow)
\end{equation}
is the normalized number density of the  observed particle in a given
kinematical region D when the projectile is polarized upwards ($\uparrow$)
/polarized downwards ($\downarrow$). $\sigma_{in}$ is the total
inelastic cross section, $x_F\equiv 2 p_{||}/\sqrt{s}$, where
$p_{||}$ is the longitudinal component of the momentum of the produced
particle  and $s$ is the total c.m. energy squared of the
incoming hadrons, $Q$ is the invariant mass of the produced particle. 
According to the proposed picture  $\Delta N(x_F,Q|s)\equiv
N(x_F,Q|s,\uparrow)-N(x_F,Q|s,\downarrow)$ is proportional to
$\Delta D(x_F,Q|s)\equiv D(x_F,Q|s,+)-D(x_F,Q|s,-)$ that is
\begin{equation}
\Delta N(x_F,Q|s)=C \,\,\Delta D(x_F,Q|s)
\end{equation}
where $D(x_F,Q|s,\pm)$ is the number density of the
produced particles  formed
through annihilation of valence quarks polarized in the same/opposite
direction as the projectile.   
$D(x_F,Q|s,\pm)$ is a convolution of three different functions.
The first one, $q_v^{\pm} (x^P)$, is the number density of the 
valence quarks 
$q_v$ polarized in the same/opposite direction as the transversely
polarized hadron, $q_s(x^T)$ is the number density of the sea-quarks in the  
unpolarized target and $K(x^P,q_v,x^T,\bar q_s,x_F,Q|s)$ is the 
probability for the direct formation. 
For baryon production it is understood that  either 
$q_v^{\pm}(x^P)$ or $q_s(x^T)$ stands for a diquark. 
The constant $C$ describes the intensity of the surface effect. 

For meson or baryon  production  we have 
\begin{equation} 
  D(x_F,M|s,\pm )=\kappa_M q_v^{\pm}(x^P)\bar q_s(x^T)
\end{equation}
where $x^P$ and $x^{T}$ are determined by momentum conservation 
$x^P\approx x_F$ and $x^T\approx x_0/x_F$ with $x_0=m^2_M/s$ 
($m_M$ is the mass of the produced particle). 
The denominator is nothing else but $2N(x_F,M|s)$ the number density of 
 observed particles in the corresponding unpolarized reaction and 
it can be written as the sum of two terms \cite{liang,lett} :
\begin{equation}
    N(x_F,M|s)=N_0(x_F,M|s)+D(x_F,M|s)
\end{equation}
Here $D(x_F,M|s)=D(x_F,M|s,+)+D(x_F,M|s,-)$ stands for the direct
formation and is meant as the sum over all 
possible direct formation processes, $D(x_F,M|s)=\sum_i D_i(x_F,M|s)$,  
$N_0(x_F,M|s)$ stands for other contributions than direct 
formation and can be extracted from the existing data. 
With this we have for meson production:
\begin{equation}
A_N(x_F,M|s)=C\kappa _M
{ \Delta q_v(x^P) \bar q_s(x^T)
\over 2[N_0(x_F,\pi|s)+D(x_F,\pi|s)]},
\end{equation}
For lepton pair production we have 
\begin{equation}
  D(x_F,Q|s,\pm)=\sum_{q} e^2_q q_V^{\pm}(x^P) \bar q_s(x^T)
\end{equation}
where $q=q_v=q_s$ is the flavor content of the quarks and 
$x^{P/T}$ are given by $x^{P/T}=[\pm x_F + \sqrt{x_F^2+(4Q^2)/s}]/2$. 

In carrying out these calculations, it is useful to note the
following: By integrating $q_v^{\pm }(x)$ over $x$,
we obtain the average number
of the valence quarks $q_v$ polarized in the same/opposite direction
as the proton. That is, \\ 
 $\Delta q_v(x) \equiv q_v^+(x)-q_v^-(x)$  
satisfies the following constraints:
\begin{equation}
\int dx \ \Delta u_v(x)=4/3,\ \ \ \ \  \ \ \
\int dx \ \Delta d_v(x)=-1/3
\end{equation}
Since the $q^{\pm}_v(x)$'s have not yet been measured,  
in order to estimate the main features of the asymmetries 
we use the following rough estimate:
\begin{equation}
\Delta u_v(x)=(2/3)u_v(x),\ \ \ \ \ \
\Delta d_v(x)=-(1/3)d_v(x),
\label{eq:uans}
\end{equation}

\section{Left-right asymmetry in inclusive hyperon production}

In  inclusive hyperon production we have three different contributions to
the cross section:

(a) Diquarks of the projectile form  hyperons with sea quarks of
the target. 

(b) Valence quarks of the projectile form  hyperons with 
sea diquarks of the target.

(c) and finally we have the non direct formation processes in the 
central rapidity region. 

It is clear that process (c) will mainly contribute in the central
rapidity region. While the processes (a) and (b) will contribute mainly
in the projectile fragmentation region and here process (a) will contribute 
at higher $x_F$ than process (b) because valence {\it diquarks} carry in
general a larger fraction of momentum than valence {\it quarks}.

Let us discuss $\Lambda$ production first. 
We have to determine the different contributions to the 
number density of the produced $\Lambda$ hyperon in unpolarized 
proton-proton collisions $N(x_F,\Lambda |s)$ : 
\begin{equation}
N(x_F,\Lambda |s)=N_0(x_F,\Lambda |s) +D^d(x_F,\Lambda |s)+ 
D^v(x_F,\Lambda |s)
\end{equation}
$D^d(x_F,\Lambda |s)$ is the contribution from
valence diquark -- sea quark annihilation and is given by: 
\begin{equation}
D^d(x_F,\Lambda |s)=\kappa_{\Lambda}^d  (ud)_v(x^P) s_s(x^T) 
\end{equation}
$D^v(x_F,\Lambda |s)$ is the contribution for valence quark -- sea diquark 
formation:
\begin{equation}
D^v(x_F,\Lambda |s)= \kappa_{\Lambda}^v \{ 
   u_v(x^P) (ds)_s(x^T) + d_v(x^P) (us)_s(x^T) \} 
\end{equation}
where for the sea diquark distribution $(qq)_s$ with 
$q=u,d,s$ we use a convolution of 
two sea quark distributions.
For the valence diquark distribution $(ud)_v$ we use the parametrization of
\cite{diquark}. 
The two constants $\kappa_{\Lambda}^d$ and $\kappa_{\Lambda}^v$   
and the contribution coming from the non direct formation processes 
are to be determined by fits to the data. 
For the functional form of the non direct formation we choose: 
$300 (1-x_F)^2 e^{-3 x_F^3}$.  
In Fig.2 a comparison is made with the data of Ref. \cite{cross}.  
The data can be reproduced very well, suggesting that hyperons 
are indeed produced by direct formation Eq.(11) and Eq.(12) 
in the fragmentation region.

In order to discuss the asymmetry in the process 
$p(\uparrow ) + p(0) \rightarrow \Lambda + X$ we note that only valence
quarks of the polarized projectile contribute to the asymmetry and we
have in detail the following behavior for the different terms: 
$u_v + (ds)_s  \rightarrow \Lambda$ contributes positively,  
$d_v + (us)_s  \rightarrow \Lambda$   negatively and 
the process $(ud)_v +s_s   \rightarrow \Lambda$ contributes also
negatively to the asymmetry.  
The signs of the first two contributions are due to 
Eq.(8). In order to understand the sign of the last contribution, 
we note the following : 
This direct formation process should be predominantly accompanied 
by the production of a meson, directly formed through 
fusion of the $u$-valence quark of the projectile 
with a suitable anti-sea-quark of the target. 
This meson should have a large probability to go left. 
Thus according to momentum
conservation, the $\Lambda$ produced through this process should go
right. This implies that this formation process contributes negatively
to the asymmetry. 
Thus these contribution is opposite in sign to that of the associatively
produced meson and is proportional to 
$- r_u(x)\equiv -\Delta u_v(x)/u_v(x)$, where
$x$ is the fractional momentum of the $u_v$ valence quark. 
Thus we have for the asymmetry:
\begin{eqnarray}
 & & A_N(x_F,\Lambda |s)=\nonumber\\
   & &    C\frac{\kappa_{\Lambda}^v [\Delta u_v(x^P)
       + \Delta d_v(x^P)] (qq)_s(x^T) -
        \kappa_{\Lambda}^d r_u(x) (ud)_v(x^P) s_s(x^T)}
       {2[N_0(x_F,\Lambda|s)+D^d(x_F,\Lambda|s)+D^v(x_F,\Lambda |s)]}
\end{eqnarray}
Since the non direct formation processes do not contribute to the 
asymmetry, it will be zero in the central rapidity region $x_F<0.2$. 
In the region where the first two processes already contribute 
$0.3<x_F<0.5$ the asymmetry will be slightly positive 
because we have $(\Delta u_v(x^P) +\Delta d_v(x^P)) 
(qq)_s(x^T)=(\frac{2}{3} u_v(x^P)-\frac{1}{3} d_v(x^P))(qq)_s(x^T)$.  
Finally in the large $x_F$ region the contribution 
$(ud)_v + s_s \rightarrow \Lambda$ dominates and the asymmetry will be 
large and negative. 
If we  use the ansatz of Eq.(10), we have for $r_u(x)=2/3$. 
In Fig.3 a comparison is made with the data using the constants 
for $\kappa_{\Lambda}^i$ determined from the unpolarized cross section. 

Since for other hyperons the unpolarized  differential cross
sections are not yet well known we can only make qualitative predictions 
for these asymmetries. 
For $\Sigma^-=dds$  the only contribution 
in the projectile fragmentation region comes from  
$d_v + (ds)_s \rightarrow \Sigma^-$ i.e. the asymmetry will be negative. 
For $\Sigma^+=uus$  we have the contribution 
 $u_v + (us)_s \rightarrow \Sigma^+$ which is positive and 
  we have also  contribution from 
$(uu)_v+s_s \rightarrow \Sigma^+$  which  contributes positively 
to the asymmetry. 
Thus $\Sigma^+$ behaves like $\pi^+$ and $\Sigma^-$ behaves like
$\pi^-$.   
Althoug $\Sigma^0=uds$ has the same flavor content as 
$\Lambda$, there is the following difference: 
While the $ud$-valence-diquark in $\Lambda $ is in
the state with the sum of
their total angular momenta $j_{ud}=0$,
the $ud$-valence-diquark
in $\Sigma ^0$ is in a $j_{ud}=1$ state.
We note that the proton wave function \cite{liang}
can be written in the following way,
\begin{eqnarray}
|p(\uparrow) \rangle &=& {1\over 2\sqrt{3}} \Bigl \{
	3u(\uparrow)\cdot {1\over \sqrt{2}}
	\Bigl [ u(\uparrow) d(\downarrow)-u(\downarrow) d(\uparrow)
\Bigr ] \nonumber\\
& & + u(\uparrow)\cdot {1\over \sqrt{2}}
\Bigl [ u(\uparrow) d(\downarrow)+u(\downarrow) d(\uparrow) \Bigr ]
-\sqrt {2} u(\downarrow) u(\uparrow) d(\uparrow) \Bigr \},
\end{eqnarray}
where $(\uparrow )$ and $(\downarrow )$ denote  the
$z$-component of the total angular momentum of the corresponding
valence quark which is  $j_z=+1/2$ or $j_z=-1/2$ respectively.
We see clearly that, if the $ud$-diquark is in a
$j_{ud}=1$-state, the other $u$-valence-quark is either
polarized upward or downward, 
with relative probabilities $1:2$.
So if $\Sigma^0 $ is produced through the same kind
of direct formation process as shown in (a),
the associatively produced meson should have a large
probability to go right.
Hence, we expect that $A_N(x_F,\Sigma^0|s)$ behaves differently
from $A_N(x_F,\Lambda|s)$ does
in the large $x_F$ region ($x_F > 0.6$, say).
In contrast to $A_N(x_F,\Lambda|s)$,
$A_N(x_F,\Sigma ^0|s)$ is positive
in sign in this region. 
This implies also that there is no change of sign in
$A_N(x_F,\Sigma ^0|s)$ as a function of  $x_F$.

For $\Xi^-=dss$ the contributions are $d_v + (ss)_s \rightarrow \Xi^-$, 
so that the asymmetry is negative while the asymmetry is positive for 
$\Xi^0=uss$  due to $u_v + (ss)_s \rightarrow \Xi^0$. 
Furthermore anti-$\Lambda$ should not show any asymmetries 
since only sea-quarks or sea-diquarks of the polarized projectile 
contribute. However for antiproton beam anti-$\Lambda$ behaves in the
same way as $\Lambda$ for proton beam, and $\Lambda$ in the same way as 
anti-$\Lambda$ for proton beams (i.e. no asymmetry).
These features are  summarized in the following table for 
proton-proton collision:

\begin{center}
\begin{tabular}{|c|c|c|c|c|c|c|c|}
\cline{1-8}
hyperon & $\Sigma^-$ & $\Sigma^0$ &$\Sigma^+$ & $\Xi^-$ & $\Xi^0$ & 
$\Lambda$ & $ anti-\Lambda$ \\
\cline{1-8} 
 asymmetry & neg. & pos. & pos. & neg. & pos. & neg. & 0.\\
\cline{1-8}  
\end{tabular} 
\end{center}  

\section{Left-right asymmetry in other processes}

Now we come to the production of Drell-Yan pairs using pion beams 
and different polarized targets:
$\pi^{\pm} +
p(\uparrow ),n(\uparrow ),D(\uparrow ) \rightarrow l\bar l + X$. 
Since here non-direct formation processes do not contribute and 
valence-valence annihilation plays the dominating role we expect that
the asymmetries are large not only in 
the fragmentation region of the polarized particles 
(in contrast to meson/hyperon production). 
Let us discuss first the difference between $\pi^+$ and $\pi^-$ using
polarized proton target. 
Since the valence-valence contribution $\bar u_v^{\pi -}$-$u_v$ plays
the dominating role in the whole kinematic region for
$\pi^- + p(\uparrow) \rightarrow l\bar l +X$  this asymmetry is
expected to be negative and its magnitude be large. 
In $\pi^+ + p(\uparrow) \rightarrow l\bar l +X$ the valence
contribution is due to $\bar d_v^{\pi^+}$-$d_v$ which is of the
same order of magnitude as the \\ $\bar u_s^{\pi +}$-$u_v$-contribution
 or even smaller in the target fragmentation region.
We expect that the corresponding asymmetry is negative in
the target fragmentation region and positive in the projectile
fragmentation region. 

If one uses the same beam, say $\pi^-$, and different projectiles, one 
expects also considerable differences. 
We compare $\pi^- +p(\uparrow ) \rightarrow l\bar l + X$ with 
$\pi^- +n(\uparrow ) \rightarrow l\bar l + X$. In both reactions 
the asymmetries are due to $\bar u_v^{\pi -}$-$u_v$. 
But because of isospin symmetry 
$\bar u_v^{\pi -}$-$u_v$ contributes in $\pi^- +p(\uparrow ) \rightarrow
l\bar l + X$  negatively to $A_N$ (note that now the target, not the 
projectile,  is polarized 
therefore the minus sign) and it contributes positively in 
$\pi^- +n(\uparrow ) \rightarrow l\bar l + X$. 
In Fig.3 we see that the asymmetry depends very much on what kind of
target we have. 
A similar discussion can be made for $\pi^+$ using different polarized
projectiles. The results should not be repeated here, they are published
in Ref. \cite{lett}.

According to the proposed picture  W/Z-boson production processes in
single-spin experiments should also show left-right asymmetries.
Since the largest contributions  come  from the
annihilation of anti-valence quarks of the polarized antiproton beam
and valence-quarks of the unpolarized proton target in the reaction
$\bar p(\uparrow ) + p(0) \rightarrow W/Z +X$ 
we only need to take them into account. Here we see that the asymmetry
for $W^+$  is negative due to $\bar d^{\bar p}_v$-$u^p_V$ and that for
$W^-$ is positive due to $\bar u^{\bar p}_v$-$d^p_V$ and that for $Z$ is
positive but smaller than that for $W^-$.

In the case of $p(\uparrow ) + p(0) \rightarrow W/Z +X$  the main
contributions to the asymmetries come from the annihilation of
valence-quarks of the polarized projectile and anti-sea quarks of the
unpolarized target. Here we have positive asymmetry for $W^+$ due to
$u^{p}_v$-$\bar d^p_s$ and negative asymmetry for $W^-$ due to
$d^{p}_v$-$\bar u^p_s$  annihilation.
$W^+$ and $W^-$ change their role in proton and antiproton induced
reactions. The sign for the asymmetry of the $Z$-boson remains the same.

\section{Conclusions} 
It is pointed out that not only  
the  asymmetries observed in inclusive meson production 
processes but also the
recent experimental data for $\Lambda$ hyperon production 
are in good agreement with the Berliner  relativistic quark model.  
Further predictions for other hyperon productions,   
Drell-Yan pair  and W/Z-boson productions have been presented.

\newpage

{\large {\bf Figure captions}}

\vspace{1.truecm}
Fig.1. The differential cross section $E d\sigma /dp^3$  
$p(0)+p(0)\rightarrow \Lambda + X$ 
at ISR-energy $\sqrt{s}=62.3$ GeV/c
and for $p_{\perp} = 0.65$ GeV/c [11]. 
The different contributions : valence diquark -- sea quark, 
valence quark -- sea diquark and non-direct formation  
 are plotted as dashed-dotted, dotted and dashed lines respectively.   
The solid line is the sum of all contributions. 
The spin averaged distribution are from Ref. 
[13], the diquark distributions are from Ref.[12].

\vspace{1.truecm}
Fig.2. The left-right asymmetry $A_N$ for $p(\uparrow)+p(0)\rightarrow 
\Lambda +X$. The data at 
$200$ GeV/c are from Ref. [15] and the low energy data 
are from Ref. [3]. The constant $C$ which parametrizes the surface
effect is chosen to be $0.6$ as in [8].

\vspace{1.truecm}
Fig.3. The left-right asymmetry $A_N$ for lepton pair production 
using $70$ $GeV/c$ $\pi^-$  beam and different polarized targets.  
The spin averaged distribution are from Ref. [13] for 
proton and from Ref. [14] for pion.

\vspace{1.truecm}
Fig.4. The left-right asymmetry $A_N$ for W/Z-boson production,  
$p(\uparrow ) + p(0) \rightarrow W^{\pm}/Z +X$ at $\sqrt{s}=500$ $GeV/c$.    
The same parametrization for the quark-distribution functions has been
used as in Fig.3.

\newpage

\begin{figure}
\psfig{file=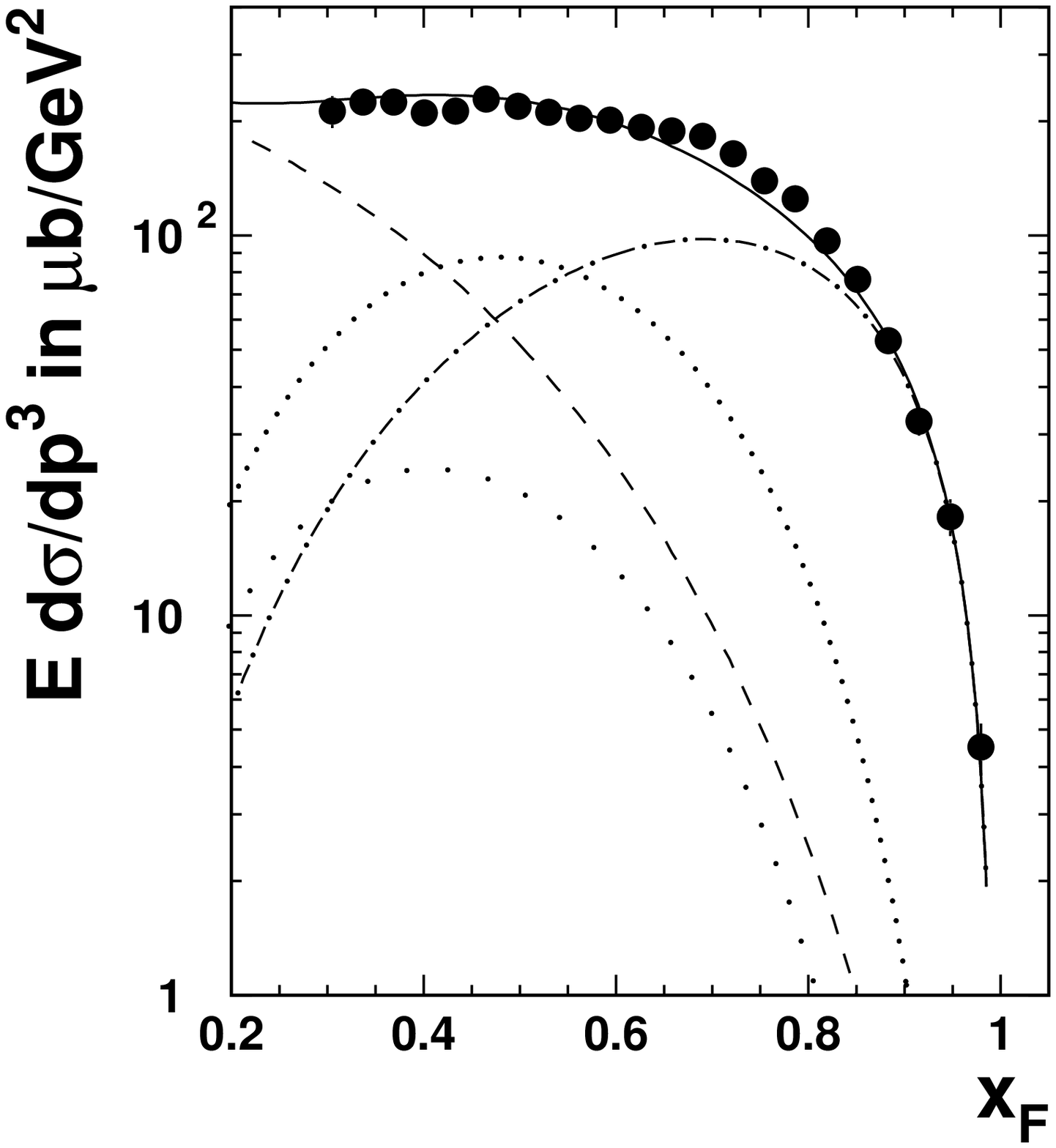,width=8.5truecm}
\caption{}
\end{figure}

\begin{figure}
\psfig{file=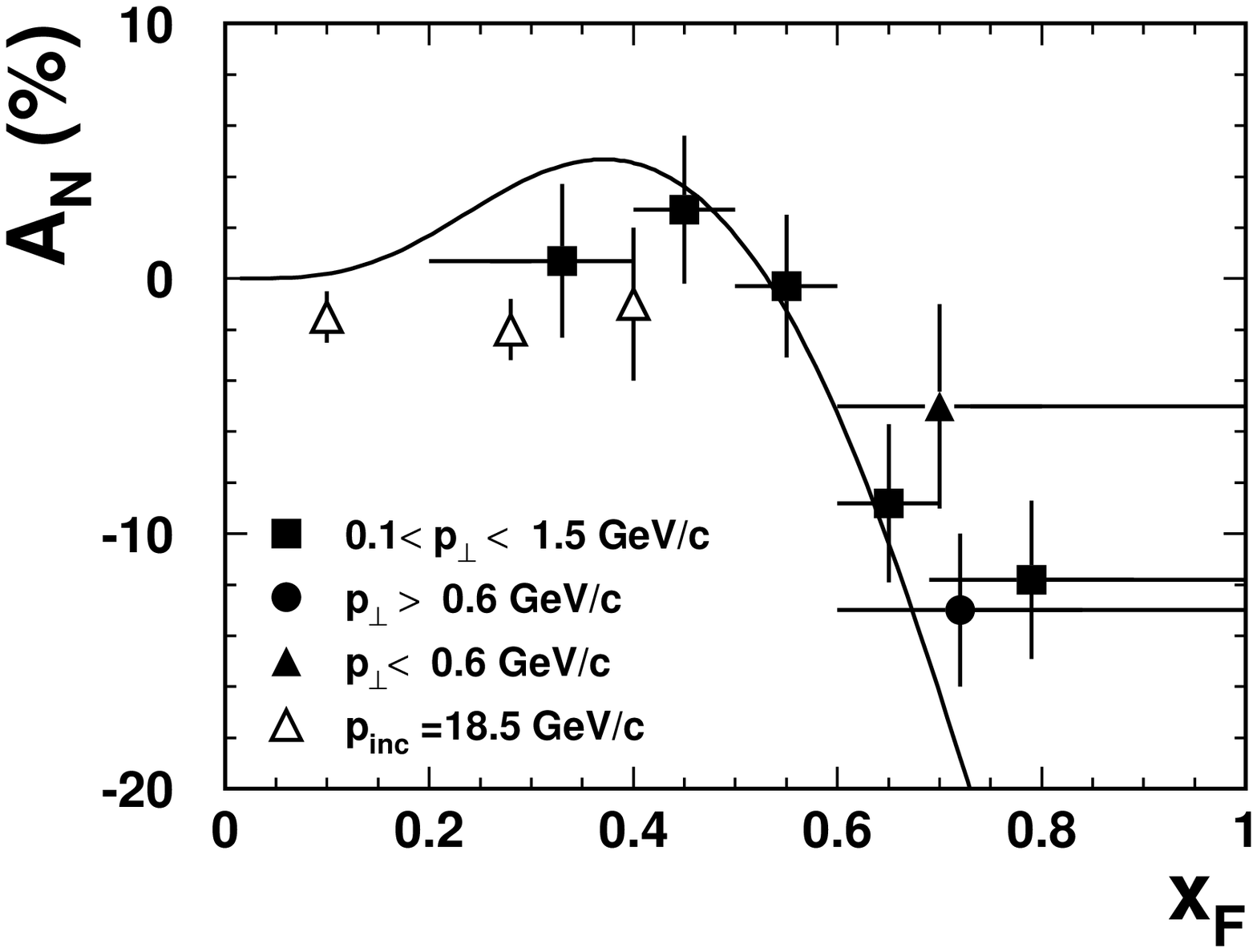,width=12truecm}
\caption{}
\end{figure}

\begin{figure}
\psfig{file=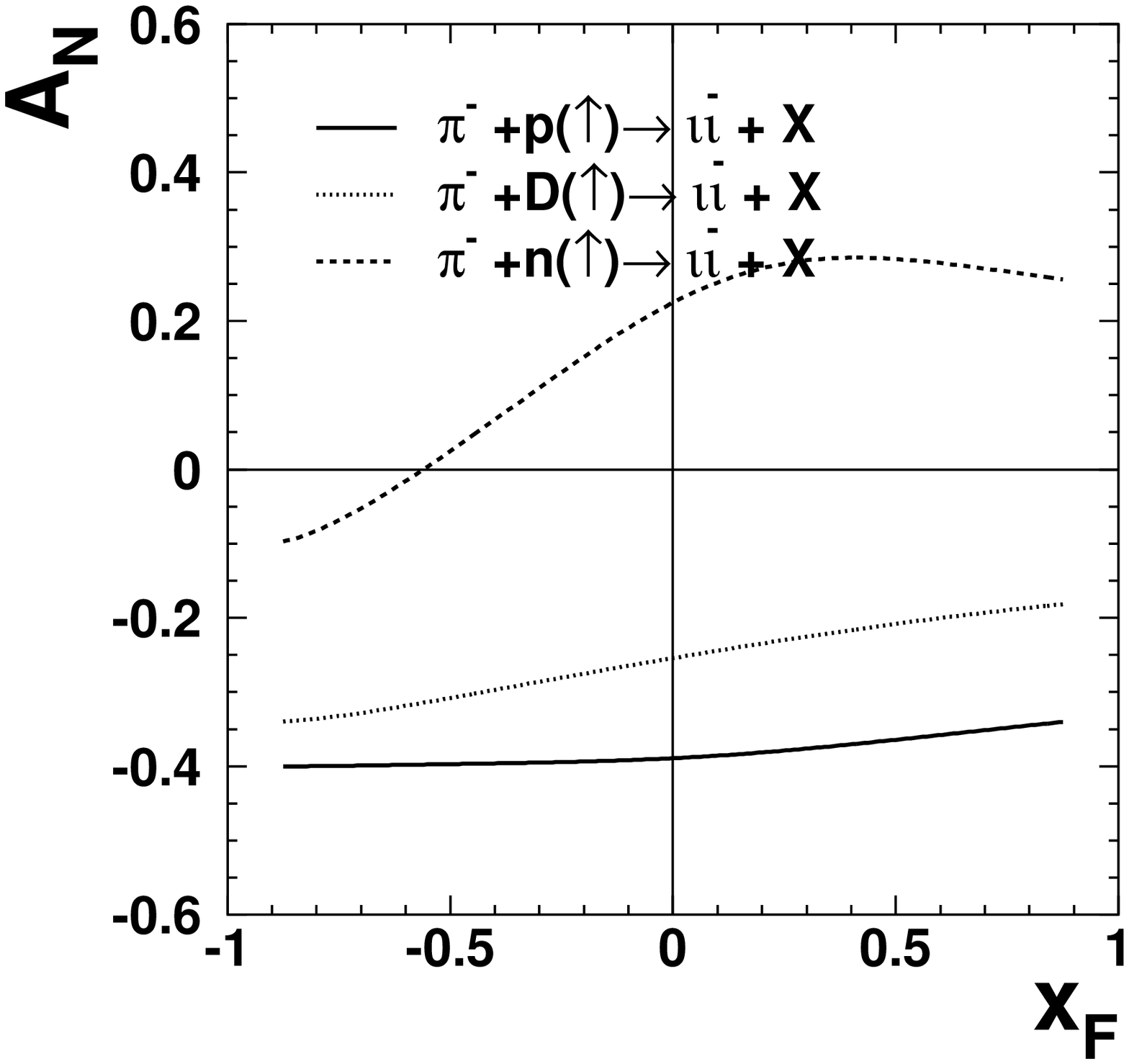,width=10truecm}
\caption{}
\end{figure}

\begin{figure}
\psfig{file=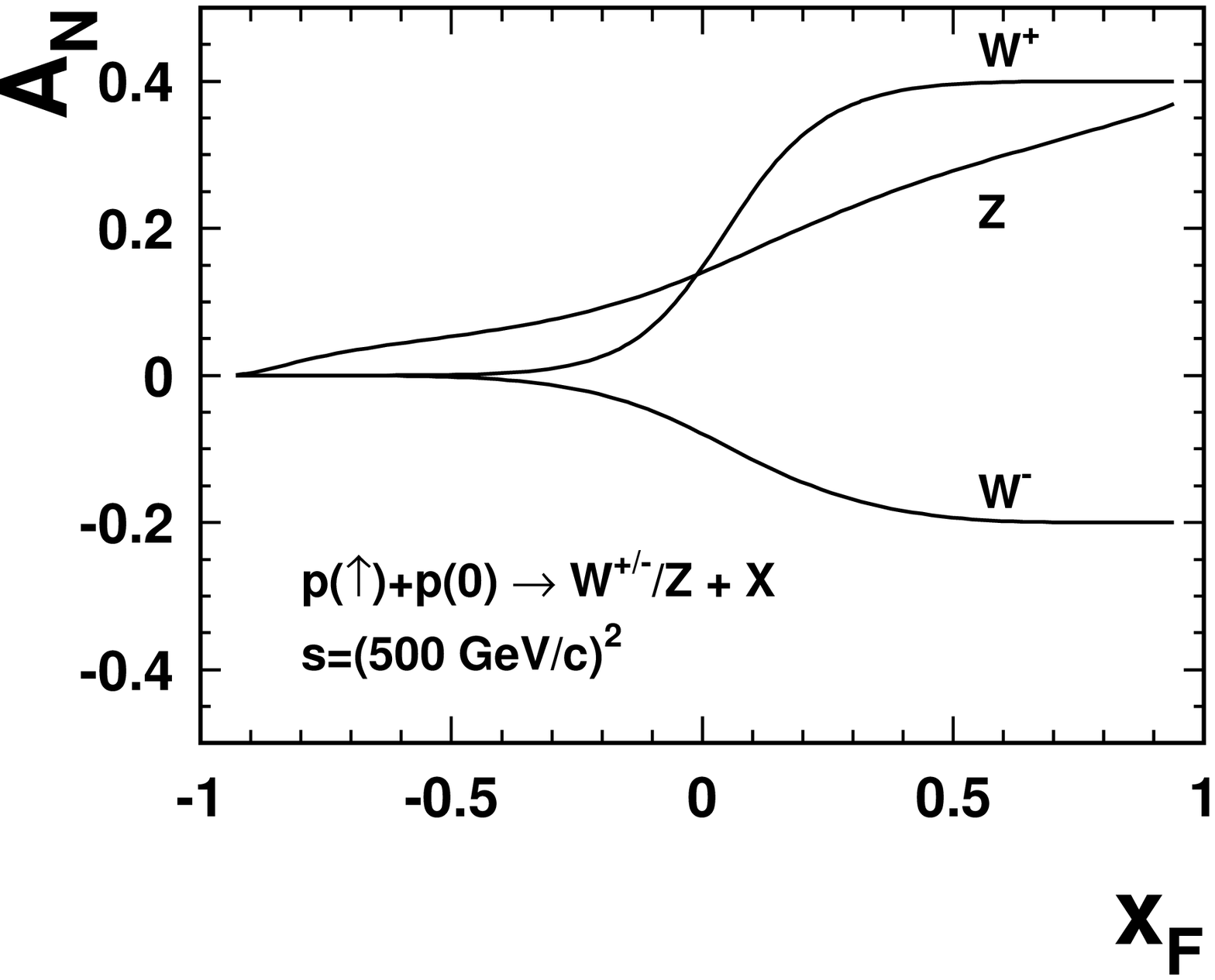,width=10truecm}
\caption{}
\end{figure}

\end{document}